\documentclass{ws-procs9x6}

\usepackage{epsfig}

\begin{document}

%
\catchline{}{}{}{}{}
%

\title{Exotic pentaquarks, crypto-heptaquarks and linear three-hadronic molecules}

\author{\footnotesize P. Bicudo}

\address{Dep. F\'{\i}sica and CFIF, Instituto Superior T\'ecnico,
Av. Rovisco Pais, 1049-001 Lisboa, Portugal,
\\ email : bicudo@ist.utl.pt}

\maketitle

\abstracts{
In this talk, multiquarks are studied microscopically in a standard quark model. 
In pure ground-state pentaquarks the short-range interaction is computed 
and it is shown to be repulsive, a narrow pentaquark cannot be in the groundstate.
As a possible excitation, an additional quark-antiquark pair is then considered, and 
this is suggested to produce linear molecular system, with a narrow decay width.
This excitation may be energetically favourable to the p-wave excitation 
suggested by the other pentaquark models. 
Here, the quarks assemble in three hadronic clusters, and the central hadron 
provides stability. The possible crypto-heptaquark hadrons with exotic 
pentaquark flavours are studied. 
}


\par
Exotic multiquarks are expected since the early works of Jaffe
\cite{Jaffe},
and the masses and decays in the SU(3) exotic anti-decuplet were first 
predicted within the chiral soliton model
\cite{Diakonov1}. 
The pentaquarks have been revived recently by several searches of the
$\Theta^+$(1540) 
\cite{Nakano,Barmin,Stepanyan,Barth,Asratyan,Kubarovsky,Airapetian,Juengst,Aleev,Bai,Abdel-Bary,Knopfle,Aslanyan,Chekanov,Pinkenburg,Troyan,Raducci,Abt}, 
first discovered at LEPS
\cite{Nakano},
and by searches
of the $\Xi^{--}$(1860)
\cite{Alt,Fischer,Price}
and of the $D^{*-}p$(3100)
\cite{H1}, 
observed respectively at NA49
\cite{Alt}
and at H1
\cite{H1}. 
Pentaquark structures
have also been studied on the lattice
\cite{Csikor,Sasaki,Chiu1,Chiu2,Mathur2,Okiharu,Ishii,Alexandrou}. 
Moreover multiquarks are favoured by the presence of several 
different flavours
\cite{Richard,Cheung,Lutz}. 
The observation of the $D^{*-}p$(3100) at H1, the 
observation of double-charmed baryons at SELEX 
\cite{SELEX},
and the future search of double-charmed baryons at COMPASS
\cite{COMPASS}
suggest that new pentaquarks with heavy quarks may be 
discovered. 

\par
In this talk it is shown that the pentaquarks cannot be in the groundstate. 
The lowest excitation consists in including a light quark-antiquark pair
in the system. This results in a heptaquark and in a linear molecular 
system. The possible crypto-heptaquark hadrons with exotic pentaquark 
flavours, with strange, charmed and bottomed quarks, are studied. 
Recently this principle was used to 
suggest that the $\Theta^+$(1540) is a 
$K \bullet  \pi \bullet  N$ molecule with binding energy of 30 MeV
\cite{Bicudo00,Llanes-Estrada,Kishimoto}, 
and the $\Xi^{--}$(1862) is a $\bar K \bullet  N \bullet  \bar K$ 
molecule with a binding energy of 60 MeV
\cite{Bicudo00,Bicudo02}.
I also suggest that the new positive parity  
scalar  $D_s(2320)$ and axial $D_{s+}(2460)$ are
$\bar K \bullet  D$ and $\bar K \bullet D^*$ multiquarks 
\cite{Bicudo01},
and that the $D^{*-} p$(3100) is consistent with a  
$D^* \bullet  \pi \bullet  N$ 
linear molecule with an energy of 15 MeV above threshold
\cite{Bicudo00,Bicudo03}. 
A systematic search of similar structures has also been
performed
\cite{Bicudo04}.
These recent results are now reviewed.


Here I study multiquarks microscopically with a standard quark-model (QM) 
Hamiltonian. The energy of the multiquark state, and the short range interaction 
of the mesonic or baryonic subclusters of the multiquark are computed 
with the multiquark matrix element of the QM Hamiltonian,
\begin{equation}
H= \sum_i T_i + \sum_{i<j} V_{ij} +\sum_{i \bar j} A_{i \bar j} \ . 
\label{Hamiltonian}
\end{equation}
Each quark or antiquark has a kinetic energy $T_i$.
The colour-dependent two-body interaction $V_{ij}$ includes 
the standard confining and hyperfine terms,
\begin{equation}
V_{ij}= \frac{-3}{16} \vec \lambda_i  \cdot   \vec \lambda_j
\left[V_{conf}(r) + V_{hyp} (r) { \vec S_i } \cdot { \vec S_j }
\right] \ .
\label{potential}
\end{equation}
The potential of eq. (\ref{potential}) 
reproduces the meson and baryon spectrum with quark and antiquark
bound states (from heavy quarkonium to the light pion mass).
Moreover, the Resonating Group Method (RGM)
\cite{Wheeler}
was applied by Ribeiro, 
\cite{Ribeiro} 
Toki
\cite{Toki} 
and Oka 
\cite{Oka}
to show that in exotic
$N + N$ scattering the quark two-body-potential, together with
the Pauli repulsion of quarks, explains the $N + N$ hard core
repulsion. 
Recently, a breakthrough was achieved in chiral symmetric quark 
models. These models are inspired in the original work of
Nambu and Jona-Lasinio
\cite{Nambu}.
Addressing a tetraquark system with $\pi+\pi$ quantum 
numbers, it was shown that the QM with the quark-antiquark 
annihilation $A_{i \bar j}$ also fully complies with chiral 
symmetry, including the Adler zero and the Weinberg theorem
\cite{Bicudo0,Bicudo1,Bicudo2}. 

\par
For the purpose of this talk, only the matrix elements of the
potentials in eq. (\ref{Hamiltonian}) matter. 
The hadron spectrum constrains the hyperfine potential,
\begin{equation}
\langle V_{hyp} \rangle \simeq \frac{4}{3} 
\left( M_\Delta-M_N \right)
\simeq M_{K^*}- M_K  \ .
\label{hyperfine}
\end{equation}
The pion mass
\cite{Bicudo3},
constrains the annihilation potential,
\begin{equation}
\langle A \rangle_{S=0} \simeq - {2 \over 3} 
(2M_N-M_\Delta)
\ ,
\label{sum rules}
\end{equation}
and this is correct for the annihilation of $u$ or $d$ quarks.

\par
The annihilation potential only shows up in non-exotic channels, and it 
is clear from eq. (\ref{sum rules}) that the annihilation potential 
provides an attractive (negative) interaction. 
The quark-quark(antiquark) potential is 
dominated by the interplay of the hyperfine interaction of eq. (\ref{hyperfine})
and the Pauli quark exchange.
In s-wave systems with low spin this results in a repulsive interaction. 
Therefore, I arrive at the attraction/repulsion criterion for groundstate hadrons:
\\
- {\em whenever the two interacting hadrons have quarks (or antiquarks)
with a common flavour, the repulsion is increased by the Pauli principle;
\\
- when the two interacting hadrons have a quark and an
antiquark with the same flavour, the attraction 
is enhanced by the quark-antiquark annihilation}.
\\
For instance, $uud-s \bar u$ is attractive,
and $uud-u \bar s$ is repulsive. 
This qualitative rule is confirmed by quantitative 
computations of the short-range interactions of the 
$\pi , \, N , \, K , \, D , \, D^* , \, B , \, B^* $ 
\cite{Bicudo00,Bicudo01,Bicudo03,Bicudo02,Bicudo0,Bicudo1,Bicudo2}.


\par
The attraction/repulsion criterion shows clearly that the exotic groundstate 
pentaquarks, containing five quarks only, are repelled. If the pentaquark could 
be forced to remain in the groundstate, this repulsion would provide a mass of 
1535 MeV, close to the $\Theta^+$ mass. There is an evidence of such a negative 
parity state both in quark model calculations and in lattice computations.
However the existence of this groundstate can only appear as an artifact in 
simulations that deny the decay into the $K-N$ channel. 
Actually the groundstate is completely open to a strong decay into the $K-N$ 
channel, and this decay is further enhanced by the repulsion. It is indeed 
well known that any narrow pentaquark must contain an excitation, to prevent a 
decay width of hundreds of MeV to a meson-baryon channel. This is understood
in the diquark and string model of Jaffe and Wiczek
\cite{Jaffe2} and 
Karliner and Lipkin 
\cite{Lipkin}, 
and in the Skyrme model of Diakonov, Petrov and Polyakov
\cite{Diakonov1}. 
These models
suggest that the pentaquarks include a p-wave, or rotational excitation. 
However this excitation usually leads to a higher energy shift than the
one observed, and a novel energy cancellation remains to be consistently
provided. A candidate for the energy cancellation is the flavour-hyperfine 
interaction of Stancu and Riska
\cite{Stancu}. 
Although these models are quite appealing, and they 
have been advocating pentaquarks for a long time, here I propose a different 
mechanism, which is more plausible in a standard quark model approach. 
Moreover this mechanism is in a sense confirmed in recent lattice computations, 
where pentaquarks with p-wave excitations indeed have a higher mass than the
observed pentaquarks. 

\par
In this talk I consider that a s-wave flavour-singlet light quark-antiquark 
pair $l \bar l$ is added to the pentaquark $M$. The resulting heptaquark 
$M'$ is a state with parity opposite to the original $M$ 
\cite{Nowak}, 
due to the intrinsic parity of fermions and anti-fermions. 
The ground-state of $M'$ is also naturally rearranged in a s-wave baryon 
and in two s-wave mesons, where the two outer hadrons are 
repelled, while the central hadron provides stability.
Because the s-wave pion is the lightest hadron, the minimum energy needed to 
create a quark-antiquark pair can be as small as 100 MeV. This energy shift 
is lower than the typical energy of 300-600 MeV of spin-isospin or angular 
excitations in hadrons. Moreover, the low-energy p-wave decay of the heptaquarks $M'$ 
(after the extra quark-antiquark pair is annihilated) results in a very narrow 
decay width, consistent with the observed exotic flavour pentaquarks.

%
%
\begin{table}[t]
\tbl{ 
Exotic-flavour pentaquarks with no heavy quark.
}
{\begin{tabular}{@{}cccc@{}}
\toprule
flavour & linear molecule   
& $~$ mass [GeV] & decay channels \\ \colrule
$ I=1/2, \ ssss \bar l (+3 \, l \bar l) $ : &five-hadron molecule
\\
\colrule
$ I=1, \ sssl \bar l  (+2 \, l \bar l) $ : & four-hadron molecule
\\
\colrule
$ I=3/2, \ ssll \bar l  (+l \bar l)  = $ & $s \bar l \bullet   lll \bullet   s \bar l :$ 
\\
& $ \bar K \bullet   N \bullet   \bar K {\bf = \Xi^{--} } $&  {\em 1.86 } &$ 
\bar K + \Sigma, \, \pi + \Xi
$\\
\colrule
$I=2, \ slll \bar l  (+l \bar l) = $ & $s\bar l \bullet   lll \bullet   l \bar l $: &pion unbound
\\
\colrule
$I=5/2, \ llll \bar l (+l \bar l) = $ & $l\bar l \bullet   lll \bullet   l \bar l$: & pion unbound
\\
\colrule
$I=0, \ llll \bar s (+l \bar l) = $ & $l \bar s \bullet   l\bar l \bullet   lll $ :
\\   
& $K \bullet   \pi \bullet   N { \bf =\Theta^+ }  $& {\em 1.54 }  &$ 
K + N 
$\\ 
\botrule
\end{tabular}}
\end{table}

\par
I now detail the strategy to find the possible linear heptaquark molecules, 
neglecting higher Fock space excitations. 
\\
{\bf a)} The top quark is excluded because it is too unstable. 
To minimise the short-range repulsion and to increase the attraction of 
the three-hadron system, I only consider pentaquarks with a minimally
exotic isospin, and with low spin. 
\\
{\bf b)} Here the flavour is decomposed in an s-wave system of 
a spin $1/2$ baryon and two 
pseudoscalar mesons, except for the vectors $D^*$ and $B^*$ 
which are also considered. 
\\
{\bf c)} I consider as candidates for narrow pentaquarks
systems where one hadron is attracted by both other ones.
The criterion is used to discriminate which hadrons
are bound and which are repelled. 
\\
{\bf d)} In the case of some exotic flavour pentaquarks, only a 
four-hadron-molecule or a five-hadron-molecule would bind. These cases 
are not detailed, because they are difficult to create in the laboratory.
%
\begin{table}[t]
\tbl{
Exotic flavour pentaquarks with one heavy quark. 
}
{\begin{tabular}{@{}cccc@{}}
\toprule
flavour& linear molecule   
& $~$ mass [GeV] & decay channels \\ \colrule
$ I=1/2, Hsss \bar l (+2 \, l \bar l) $ : & four-hadron molecule
\\ 
\colrule
$ I=1, Hssl \bar l (+l \bar l)= $ & $     s \bar l \bullet  l l H \bullet  s \bar l   :$ 
\\
&$ \bar K \bullet  \Lambda_c \bullet  \bar K $&$  3.23 \pm 0.03 $&$  \bar K + \Xi_c  , \, \pi +\Omega_c 
$\\
&$ \bar K \bullet  \Lambda_b \bullet  \bar K $&$  6.57 \pm 0.03 $&$  \bar K + \Xi_b  , \, \pi +\Omega_b
$\\
\colrule
$ I=3/2, Hsll \bar l (+l \bar l)=  $ & $    s \bar l \bullet  lll \bullet  H \bar l   :$
\\
&$  \bar K \bullet  N \bullet  D $&$ 3.25 \pm 0.03 $&$ \bar K + \Sigma_c  , \, D + \Sigma , \, \pi + \Xi_c 
$\\ 
&$ \bar K \bullet  N \bullet  D^* $&$  3.39 \pm 0.03 $&$  \bar K + \Sigma_c  , \, D^* + \Sigma , \, \pi + \Xi_c 
$\\ 
&$ \bar K \bullet  N \bullet  \bar B $&$ 6.66 \pm 0.03 $&$ \bar K + \Sigma_b  , \, \bar B + \Sigma , \, \pi + \Xi_b

$\\ 
&$ \bar K \bullet  N \bullet  \bar B^* $&$  6.71 \pm 0.03 $&$  \bar K + \Sigma_b  , \, \bar B^* + \Sigma , \, \pi + \Xi_b

$\\
\colrule
$ I=2, Hlll \bar l (+l \bar l)= $ & $ l \bar l \bullet  l l l \bullet  H \bar l   $ 
:  & pion unbound
\\
\colrule  
$ I=1/2, Hlll \bar s  (+l \bar l)= $ & $ l \bar s \bullet  l \bar l \bullet  l l H  :$ 
\\  
&$ K \bullet  \pi \bullet  \Sigma_c $&$ 3.08 \pm 0.03 $&$  K + \Lambda_c , \, K + \Sigma_c  , \, D_s +N
$\\
&$ K \bullet  \pi \bullet  \Sigma_b $&$ 6.41 \pm 0.1  $&$  K + \Lambda_b , \, K + \Sigma_b  , \, D_s +N
$\\
$ I=1/2, Hlll \bar s  (+l \bar l) = $ & $ l \bar s \bullet  H \bar l \bullet  lll  :$ 
\\  
&$ K \bullet  \bar D \bullet  N $&$  3.25 \pm 0.03 $&$  K + \Lambda_c , \ K + \Sigma_c , \ D_s + N 
$\\
&$ K \bullet  \bar D^* \bullet  N $&$ 3.39  \pm 0.03 $&$  K + \Lambda_c  , \ K + \Sigma_c  , \ D_s^* + N 
$\\
&$ K \bullet  \bar B \bullet  N $&$  6.66 \pm 0.03 $&$  K + \Lambda_b  , \ K + \Sigma_b , \ B_s + N  
$\\
&$ K \bullet  \bar B^* \bullet  N $&$ 6.71 \pm 0.03 $&$   K + \Lambda_b  , \ K + \Sigma_b , \ B_s^* + N 
$\\
\botrule
\end{tabular}}
\end{table}
\\
{\bf e)} Moreover, in the particular case where one of the three hadrons 
is a $\pi$, binding is only assumed if the $\pi$ is the central hadron, 
attracted both by the other two ones. The $\pi$ is too light 
to be bound by just one hadron
\cite{Bicudo00}.
\\
{\bf f)} The masses of the bound states with a pion are computed assuming a total 
binding energy of the order of 10 MeV, averaging the binding energy of the
$\Theta^+$ and of the $D^{*-}p$ system in the molecular perspective. The masses of the 
other bound states are computed assuming a total binding energy of the order of 50 MeV, 
averaging the binding energies of the $\Xi^{--}$ and of the new positive-parity $D_S$
mesons. 
\\
{\bf g)}
This results in an error bar of $\pm$ 30 MeV for the mass.
When one of the three hadrons is not listed by the Particle Data Group
\cite{RPP}, 
its mass is extracted from a lattice computation
\cite{Mathur},
and the error bar is $\pm$ 100 MeV.
\\
{\bf e)}
Although three-body decay channels are possible through quark rearrangement,
their observation requires high experimental statistics. Only some of the 
different possible two-body decay processes are detailed here.


\begin{table}[t]
\tbl{
Exotic flavour pentaquarks with one heavy anti-quark.
}
{\begin{tabular}{@{}cccc@{}}
\toprule
flavour& linear molecule   
& $~$ mass [GeV] & decay channels \\ \colrule
$ I=0, ssss \bar H (+3 l \bar l)$ : & five-hadron molecule
\\
\colrule
$ I=1/2, sssl \bar H (+2 \, l \bar l)  $ : & four-hadron molecule
\\
\colrule
$ I=0, ssll \bar H (+l \bar l) = $ & $ l \bar H \bullet  l \bar l \bullet  lss $ 
\\
&$  \bar D \bullet  \pi \bullet  \Xi $&$  3.31 \pm 0.03 $&$  \bar D + \Xi  
, \, \bar D_s + \Lambda 
$\\
&$  \bar D^* \bullet  \pi \bullet  \Xi $&$  3.45 \pm 0.03 $&$  \bar D^* + \Xi
, \, \bar D^*_s + \Lambda , \, \bar D_s + \Lambda 
$\\
&$  B \bullet  \pi \bullet  \Xi $&$  6.73 \pm 0.03 $&$  B + \Xi  
, \, B_s + \Lambda
$\\
&$  B^* \bullet  \pi \bullet  \Xi $&$  6.77 \pm 0.03 $&$  B^* + \Xi  
, \, B^*_s + \Lambda , \, B_s + \Lambda
$\\
\colrule
$ I=1/2, slll \bar H (+l \bar l) = $ & $ l \bar H \bullet  l \bar l \bullet  l l s  $ 
\\
&$ \bar D \bullet  \pi \bullet  \Sigma $&$  3.19 \pm 0.03 $&$  \bar D + \Lambda , \, \bar D + \Sigma  , \, \bar D_s +N
$\\
&$ \bar D^* \bullet  \pi \bullet  \Sigma $&$  3.33 \pm 0.03 $&$  \bar D^* + \Lambda , \, \bar D^* + \Sigma  , \, \bar D^*_s +N
$\\
&$ B \bullet  \pi \bullet  \Sigma $&$  6.60 \pm 0.03 $&$  B + \Lambda , \, B + \Sigma  , \, B_s +N 
$\\
&$ B^* \bullet  \pi \bullet  \Sigma $&$  6.64 \pm 0.03 $&$  B^* + \Lambda , \, B^* + \Sigma  , \, B^*_s +N
$\\
$ I=1/2, slll \bar H (+l \bar l) =   $ & $ l \bar H \bullet  s\bar l \bullet  l l l   $ 
\\
&$  \bar D \bullet  \bar K \bullet  N $&$  3.25 \pm 0.03 $&$  \bar D + \Lambda , \, \bar D + \Sigma  , \, \bar D_s +N
$\\
&$ \bar D^* \bullet  \bar K \bullet  N $&$ 3.39 \pm 0.03 $&$   \bar D^* + \Lambda , \, \bar D^* + \Sigma  , \, \bar D^*_s +N
$\\
&$ B \bullet  \bar K \bullet  N $&$ 6.66 \pm 0.03 $&$  B + \Lambda , \, B + \Sigma  , \, B_s +N
$\\
&$ B^* \bullet  \bar K \bullet  N $&$  6.71 \pm 0.03 $&$  B^* + \Lambda , \, B^* + \Sigma  , \, B^*_s +N
$\\
\colrule
$ I=0, llll \bar H (+l \bar l)= $ & $ l \bar H \bullet  l \bar l \bullet  lll $
\\
& $ \bar D \bullet  \pi \bullet  N  $&$  2.93 \pm 0.03 $&$ \bar D+ N  
$\\ 
& $ \bar D^* \bullet  \pi \bullet  N {\bf  = \bar D^{*-}p } $& {\em 3.10 } &$ \bar D^*+ N , \, \bar D +N
$\\ 
& $ B \bullet  \pi \bullet  N $&$ 6.35 \pm 0.03 $&$ B+N  
$\\ 
& $ B^* \bullet  \pi \bullet  N $&$  6.39 \pm 0.03 $&$ B^*+N, B+N 
$\\ 
\botrule
\end{tabular}}
\end{table}

\par
To conclude, this work has performed a systematic search of exotic-flavour pentaquarks, 
using the heptaquark, or linear three-body hadronic-molecule perspective. 
This perspective is the result of standard QM computations
of pentaquarks and hepatquark masses and of hadron-hadron short-range interactions. 
A large number of new exotic flavour-pentaquarks are predicted in Tables
1, 2 and 3 together with their two-body decay channels. 
The systems with more than one heavy antiquark are very numerous and they are omitted 
here. 
Moreover, some new multiquarks may be easier to bind than the presently observed 
exotic pentaquarks.

\section*{Acknowledgments}
I thank the organisers of Pentaquark04, and I am grateful to 
Chris Hanhart, Eulogio Oset, Dimitri Diakonov, Frank Lee, 
Fumiko Okiharu, Hiroshi Toki, Makoto Oka, Silvia Nicolai, 
Takashi Nakano and Ting-Wai Chiu for lively discussons during 
the Pentaquark04 conference.


\end{document}